\documentclass[a4paper,10pt,twoside,bibnote]{cpc-hepnp}
\usepackage{CJK,upgreek,fancyhdr}
\usepackage{multicol}
\usepackage{graphicx}
\usepackage{booktabs}
\usepackage{slashed}
\usepackage{amssymb,bm,mathrsfs,bbm,amscd}
\usepackage[tbtags]{amsmath}
\usepackage{lastpage}
\usepackage{multirow}
\usepackage{textcomp}
\usepackage{epstopdf}
\usepackage{float}
\usepackage{footnote} 

\let\jnfont=\rm
\def\NPB#1,{{\jnfont Nucl.\ Phys.\ B, }{\bf #1}:}
\def\PLB#1,{{\jnfont Phys.\ Lett.\ B, }{\bf #1}:}
\def\EPJC#1,{{\jnfont Eur.\ Phys.\ Jour.\ C, }{\bf #1}:}
\def\PRD#1,{{\jnfont Phys.\ Rev.\ D, }{\bf #1}:}
\def\PRL#1,{{\jnfont Phys.\ Rev.\ Lett.,\ }{\bf #1}:}
\def\MPLA#1,{{\jnfont Mod.\ Phys.\ Lett.\ A, }{\bf #1}:}
\def\JPG#1,{{\jnfont J.\ Phys.\ G, }{\bf #1}:}
\def\CTP#1,{{\jnfont Commun.\ Theor.\ Phys.,\ }{\bf #1}:}
\def\ZPC#1,{{\jnfont Z.\ Phys.,\ C }{\bf #1}:}
\def\JHEP#1,{{\jnfont JHEP, \ }{\bf #1}:}
\def\lsim{\raise0.3ex\hbox{$<$\kern-0.75em\raise-1.1ex\hbox{$\sim$}}}
\def\gsim{\raise0.3ex\hbox{$>$\kern-0.75em\raise-1.1ex\hbox{$\sim$}}}

\newcommand{\GeV}{~{\rm GeV}}
\newcommand{\TeV}{~\rm TeV}

\begin{document}

\setlength{\abovecaptionskip}{4pt plus1pt minus1pt}   
\setlength{\belowcaptionskip}{4pt plus1pt minus1pt}   
\setlength{\abovedisplayskip}{6pt plus1pt minus1pt}   
\setlength{\belowdisplayskip}{6pt plus1pt minus1pt}  
\addtolength{\thinmuskip}{-1mu}            
\addtolength{\medmuskip}{-2mu}             
\addtolength{\thickmuskip}{-2mu}          
\setlength{\belowrulesep}{0pt}        
\setlength{\aboverulesep}{0pt}         
\setlength{\arraycolsep}{2pt}

\fancyhead[c]{\small Chinese Physics C~~~Vol. 42, No. 10 (2018) 103109}
\fancyfoot[C]{\small 103109-\thepage}
\footnotetext[0]{Received 11 June 2018, Published online 12 September 2018 }

\title{The semi-constrained NMSSM in light of muon g-2, LHC, and\\ dark matter constraints\thanks{Supported by
National Natural Science Foundation of China (NNSFC) (11605123,
11675147, 11547103, 11547310),  the Innovation Talent project of
Henan Province (15HASTIT017), and the Young Core Instructor
Foundation of Henan Education Department. J. Z. also thanks the
support of the China Scholarship Council (CSC) (201706275160)
while at the University of Chicago
as a visiting scholar, and the U.S. National
Science Foundation (NSF) (PHY-0855561) while at Michigan State University from 2014-2015. }}

\author{
      Kun Wang$^{1}$%
\quad Fei Wang$^{2}$%
\quad Jingya Zhu$^{1;1)}$\email{zhujy@itp.ac.cn}%
\quad Quanlin Jie$^{1}$%
}
\maketitle

\address{
$^1$ Center for Theoretical Physics, School of Physics and Technology, Wuhan University, Wuhan 430072, China \\
$^2$ School of Physics, Zhengzhou University, Zhengzhou 450000, China\\
}

\begin{abstract}
The semi-constrained NMSSM (scNMSSM) extends the MSSM by a singlet field, and requires unification of the soft SUSY breaking terms in the squark and slepton sectors, while it allows that in the Higgs sector to be different.
We try to interpret the muon g-2 in the scNMSSM, under the constraints of 125 GeV Higgs data, B physics, searches for low and high mass resonances, searches for SUSY particles at the LHC, dark matter relic density
by WMAP/Planck, and direct searches for dark matter by LUX, XENON1T, and PandaX-II.
We find that under the above constraints, the scNMSSM can still
(i) satisfy muon g-2 at 1$\sigma$ level, with a light muon sneutrino and light chargino;
(ii) predict a highly-singlet-dominated 95~GeV Higgs,  with a diphoton rate as hinted at by CMS data, because of a light higgsino-like chargino and moderate $\lambda$;
(iii) get low fine tuning from the GUT scale with small $\mu_{\rm eff},\, M_0,\, M_{1/2},\, {\rm and}\, A_0$, with a lighter stop mass which can be as low as about 500 GeV, which can be further checked
in future studies with search results from the 13~TeV LHC;
(iv) have the lightest neutralino be singlino-dominated or higgsino-dominated, while the bino and wino are heavier because of high gluino bounds at the LHC and universal gaugino conditions at the GUT scale;
(v) satisfy all the above constraints, although it is not easy for the lightest neutralino, as the only dark matter candidate, to get enough relic density. Several ways to increase relic density are discussed.
\end{abstract}


\begin{pacs}
11.30 Pb     \qquad     {\bf DOI:} 10.1088/1674-1137/42/10/103109
\end{pacs}

\footnotetext[0]{\hspace*{-3mm}\raisebox{0.3ex}{$\scriptstyle\copyright$}2018
Chinese Physical Society and the Institute of High Energy Physics
of the Chinese Academy of Sciences and the Institute
of Modern Physics of the Chinese Academy of Sciences and IOP Publishing Ltd}%

\vspace{0.7mm}
\begin{multicols}{2}

\section{Introduction}

In July 2012, the Higgs boson was discovered at the LHC
\cite{1207-a-com,1207-c-com}, and searching for physics beyond
the Standard Model (SM) has now become the main objective in high energy
physics. Supersymmetry (SUSY) is one of the most popular theories for new physics. As the simplest SUSY model, the minimal supergravity
model (mSUGRA) has attracted a lot of attention from both theorists and
experimentalists. However, it cannot predict a 125 GeV SM-like Higgs when
considering all the constraints, including muon g-2 at 2$\sigma$
level, and dark matter \cite{125-CMSSM, 125-susy}. When we give up
uniform parameters at the grand unification (GUT) scale, the MSSM can satisfy all the
constraints well, but there is a problem with fine-tuning
\cite{125-susy}.

After the Higgs boson was discovered, it became necessary to ask whether there is a second Higgs-like
particle. Searches at LEP, the Tevatron, and the
LHC have excluded a lighter SM-like Higgs, while a lighter second Higgs
with rates lower than the SM-like one could still be possible.
Recently, the CMS collaboration presented their searches for
low-mass new resonances decaying to two photons. For both the 8
TeV and 13 TeV dataset, a small excess around 95 GeV was hinted at,
with approximately 2.8$\sigma$ local (1.3$\sigma$ global)
significance for a hypothetical mass of 95.3 GeV in combined
analysis \cite{CMS95}. This result has been interpreted or discussed
in several papers \cite{interpret95}. The MSSM cannot predict such a
lighter second Higgs together with a 125 GeV SM-like Higgs under
other constraints like the muon g-2 and dark matter \cite{125-susy}.

The next-to-minimal supersymmetric Standard Model (NMSSM) has more
freedom to predict a SM-like 125 GeV Higgs, under all the
constraints and with low fine-tuning \cite{125-susy}. At the same
time, it can also predict a lighter second Higgs with rates lower
than the SM-like one  \cite{low-NMSSM, low-NMSSM98}. Since simple
models are usually more favoured, the fully constrained NMSSM
(cNMSSM) \cite{125-cNMSSM, cNMSSM-pre125, nmspec} and the
semi-constrained NMSSM (scNMSSM) are also being studied
\cite{scNMSSM-1405, 125-scNMSSM}. For the full cNMSSM,
with all soft SUSY breaking terms unified at the
GUT scale, including $M_{H_{u}}=M_{H_{d}}=M_S=M_0$, there should be
only four continuous parameters, the same as mSUGRA. While in many
studies of the cNMSSM \cite{125-cNMSSM, cNMSSM-pre125, nmspec},
there is an additional parameter $\lambda$, for a singlet scalar
$M_S$ does not in fact need to be unified. Such an issue was
also pointed out in Ref.~\cite{CNMSSM-4or5}. In the 5-parameter and 4-parameter cNMSSM,
the SM-like Higgs cannot get to 125 GeV under all the constraints
including muon g-2 \cite{125-cNMSSM,cNMSSM-pre125}.
The scNMSSM is also called the non-universal Higgs
mass (NUHM) version of the NMSSM, for it allows the soft SUSY breaking
terms in the Higgs sector to be different. In Ref.~\cite{125-cNMSSM},
the parameter $\lambda$ is always less than 0.1, so the results in
Higgs sector may not be much different from the NUHM version of
MSSM, e.g., the 125 GeV SM-like Higgs is always the lightest
Higgs. In Refs.~\cite{125-scNMSSM, scNMSSM-1405}, the   muon
g-2 constraint is set aside. In this paper, we consider all the constraints,
including muon g-2, and also require a lighter Higgs with rates
constrained by LEP, Tevatron, and LHC searches. For the dark
matter relic density, we only apply the upper bound
\cite{low-NMSSM98}, considering that there may be other sources of
dark matter \cite{dm-others}. We focus on the muon g-2,
its relation to model parameters, SUSY particle masses, and other
constraints like the dark matter relic density.

This paper is organized as follows. First, we briefly introduce  the
NMSSM and scNMSSM in Section 2. In Section 3, we discuss the
constraints on the model, present our numerical results and have
some discussion. Finally, we draw our conclusions in Section 4.

\section{The NMSSM and scNMSSM}

In the NMSSM, the Higgs sector consists of two complex doublet
superfields $\hat{H}_{u}$ and $\hat{H_{d}}$, and one complex singlet
superfield $\hat{S}$. Then the superpotential of the NMSSM with $Z_3$
symmetry is given by \cite{NMSSM-rev}
\begin{equation}
\label{eq:w_nmssm} W_{\rm NMSSM} = \lambda \hat{S} \hat{H}_{u} \cdot
\hat{H}_{d} + \frac{\kappa}{3} \hat{S} ^3 + W_{F} \,,
\end{equation}
where $W_{F}$ is the superpotential of the MSSM without the
$\mu$-term, which is the Yukawa couplings of $\hat{H}_{u}$ and
$\hat{H}_{d}$ to the quark and lepton superfields
\cite{MSSM-Martin}. At electroweak symmetry breaking, the
Higgs fields $\hat{H}_{u}\,$, $\hat{H}_{d}$ and $\hat{S}$ get their
vacuum expectation values (VEVs) $v_{u}\,$, $v_{d}$ and $v_{s}$
respectively, with $\tan\beta\equiv v_{u}/v_{d}$. Then their scalar
component fields can be written as
\begin{align}
\label{eq:higgsfields}
 H_{u} =& \left(
\begin{array}{c}
H^+_{u} \\
v_{u}+\dfrac{ \phi_{u} + i \varphi_{u}}{ \sqrt{2} }
\end{array} \right)\,, \quad
H_{d} =   \left(
\begin{array}{c}
v_{d}+\dfrac{\phi_{d} + i \varphi_{d} }{ \sqrt{2} }\nonumber\\
H^-_{d}
\end{array} \right)\,, \\
 S =&  v_{s} + \dfrac{\phi_{s}+i \varphi_{s}}{ \sqrt{2} } \,,
\end{align}
where $H^+_{i}$, $\phi_{i}$ and $\varphi_{i}$ $(i=u,d)$ represent the
charged, neutral CP-even and neutral CP-odd component fields
respectively. So the first term in $W_{\rm NMSSM}$ generates an
effective $\mu$-term,
\begin{equation}
\label{eq:mueff} \mu_{\rm eff} = \lambda v_{s} .
\end{equation}

With the superpotential, we can get the so-called `F-term' of the
Lagrangian \cite{NMSSM-rev},
\begin{equation}
\label{eq:f-term} -\mathcal{L}_{F} = \sum_{i} \Big|\frac{\delta W_{\rm
NMSSM}}{\delta \hat{\varPhi}_{i}} \Big|^2,
\end{equation}
where $\hat{\varPhi}_{i}$ can be any chiral superfield in the
superpotential.

Since the singlet field is not included, the D-term is the same as
in the MSSM,
\begin{align}
\label{eq:d-term}
-\mathcal{L}_{D} =& \frac{1}{2}g_2^2 \left(|H_{u}|^2 |H_{d}|^2 -|H_{u}\cdot H_{d}|^2\right)\nonumber\\
& + \frac{1}{8}\left(g_1^2+g_2^2\right)\left(|H_{u}|^2 - |H_{d}|^2\right)^2 .
\end{align}

The soft breaking terms in the NMSSM have 4 main parts. In the
Higgs sector, the soft breaking terms are given by
\begin{align}
\label{eq:higgssoft} -\mathcal{L}_{\rm Higgs} =&  M_{H_{u}}^2 |H_{u}|^2 +
M_{H_{d}}^2 |H_{d}|^2 + M_S^2 |S|^2 \nonumber\\
&
 + \left( \lambda A_\lambda S H_{u} \cdot H_{d} + \frac{1}{3} \kappa A_\kappa S^3 +\rm h.c. \right) \,,
\end{align}
where $M_{H_{u}}\,$, $M_{H_{d}}$\,, $M_S\,$, $A_ \lambda$ and
$A_\kappa$ are soft breaking parameters. In the mass terms of
squarks $\{\,
\tilde{q}_{i}  \equiv (\tilde{u}_{iL},\tilde{d}_{iL}),\,
\tilde{u}_{i}^c,\, \tilde{d}_{i}^c \,\}$ and sleptons $\{\,
\tilde{\ell}_{i} \equiv (\tilde{\nu}_{iL},\tilde{e}_{iL}),\,
\tilde{e}_{i}^c\,\}$ ($i = 1,2,3$ refers to generation):
\begin{equation}
\label{eq:sfermionmass} -\mathcal{L}_0  =
M^2_{\tilde{q}_{i}}|\tilde{q}_{i}|^2+M^2_{\tilde{u}_{i}}|\tilde{u}_{i}^c|^2+M^2_{\tilde{d}_{i}}|\tilde{d}_{i}^c|^2+
M^2_{\tilde{\ell}_{i}}|\tilde{\ell}_{i}|^2+M^2_{\tilde{e}_{i}}|\tilde{e}_{i}^c|^2
\,.
\end{equation}
In the mass terms of the gauginos $\tilde{B}$ (bino),
$\tilde{W}^a$ (winos) and $\tilde{G}^a$ (gluinos):
\begin{equation}
\label{eq:gaugino1} -\mathcal{L}_{1/2}  = \frac{1}{2}\left[
M_1\tilde{B}\tilde{B}
+M_2\sum\limits^3_{a=1}\tilde{W}^a\tilde{W}_a
+M_3\sum\limits^8_{a=1}\tilde{G}^a\tilde{G}_a \right] + \rm h.c. \,.
\end{equation}
In the trilinear interactions between the third generation squarks
or sleptons and the Higgs field (the Yukawa coupling of the first
two generations can be neglected):
\begin{equation}
\label{eq:gaugino2} -\mathcal{L}_{3}  = \left( h_{t} A_{t} Q \cdot H_{u}
\, \tilde{u}_3^c + h_{b} A_{b} H_{d} \cdot Q \, \tilde{d}_3^c + h_\tau
A_\tau H_{d} \cdot L \,\tilde{e}_3^c \right)+\rm h.c. \,.
\end{equation}

\textbf{Higgs sector: } In order to present the mass matrices of
the Higgs fields in a physical way, we rotate the Higgs fields by
\cite{NMSSM-mass}:
\begin{equation}
\label{eq:hdefine} H_1 = \cos\beta\, H_{u} - \epsilon \sin\beta\,
H_{d}^* \,, ~~ H_2 = \sin\beta\, H_{u} + \epsilon \cos\beta\, H_{d}^*
\,, ~~ H_3 = S \,,
\end{equation}
where $\epsilon_{12}= - \epsilon_{21} = 1$ and $ \epsilon_{11} =
\epsilon_{22} =0 \,$. With this rotation, $H_{i} \, (i=1,2,3)$ are
given by
\begin{align}
\label{eq:hfield}
H_1 =& \left(
\begin{array}{c}
H^+ \\
\dfrac{S_1 + {\rm i} P_1}{ \sqrt{2} }
\end{array} \right)\,, \nonumber\\
H_2 =& \left(
\begin{array}{c}
G^+\\
v + \dfrac{S_2 + {\rm i} G^0 }{ \sqrt{2} }
\end{array} \right)\,, \\
H_3 =&  v_{s} + \dfrac{1}{ \sqrt{2} } (S_3 + {\rm i} P_2)\,,\nonumber
\end{align}
where $G^+$ and $G^0$ are Goldstone bosons eaten by $W^+$ and $Z$
respectively, and $v=\sqrt{v_{u}^2 + v_{d}^2}=174\GeV$ is the VEV of the
Higgs field in the SM. Thus the field $H_2$ is the SM Higgs field.

In the CP-conserving NMSSM, the field $S_1\,$, $S_2$ and $S_3$ mix
to form the three physical CP-even Higgs bosons $h_{i}(i=1,2,3)$, and the
fields $P_1$ and $P_2$ mix to form the two physical CP-odd Higgs
bosons $a_{i}(i=1,2)$.

In the basis $\{S_1\,, S_2\,, S_3\}$, the elements of the
corresponding mass matrix are given by \cite{NMSSM-mass}:
\begin{eqnarray}
\label{eq:m}
M_{11}^2 &=& M_A^2 + \left(m_Z^2- \lambda^2 v^2 \right)\sin^2 2 \beta \,,\nonumber\\[2mm]
M_{12}^2 &=& - \frac{1}{2}\left(m_Z^2- \lambda^2 v^2\right) \sin 4 \beta \,,\nonumber\\[2mm]
M_{13}^2 &=& -\left( \frac{M_A^2}{2 \mu /\sin 2 \beta } + \kappa v_{s} \right) \lambda v \cos 2 \beta \,, \nonumber\\[2mm]
M_{22}^2 &=& m_Z^2 \cos^2 2 \beta + \lambda^2 v^2 \sin^2 2 \beta \,, \nonumber\\[2mm]
M_{23}^2 &=& 2 \lambda \mu v \left[1-\left( \frac{M_A}{2\mu / \sin 2 \beta} \right)^2 - \frac{\kappa}{2 \lambda } \sin 2 \beta  \right] \,, \nonumber\\[2mm]
M_{33}^2 &=& \frac{1}{4} \lambda^2 v^2 \left(\frac{M_A}{\mu / \sin 2 \beta }\right)^2 + \kappa v_{s} A_\kappa +4(\kappa v_{s})^2 \nonumber\\
          &&   - \frac{1}{2} \lambda \kappa v^2 \sin 2 \beta \,.
\end{eqnarray}
where $M_A$ is the mass scale of the doublet field $H_1$\,, and
it is given by
\begin{equation}
\label{eq:ma} M_A^2= \frac{2\mu}{\sin 2 \beta} \left( A_\lambda
+\kappa v_{s} \right) \, .
\end{equation}

The mass matrix in Eq.\eqref{eq:m} can be diagonalized by an
orthogonal matrix $[S_{ij}]$\,. We can get the mass eigenstates of
CP-even states $h_{i}$ as
\begin{equation}
\label{eq:h} h_{i} = \sum_{j=1}^3 S_{ij} S_j \, ,
\end{equation}
where $S_{ij}$ are the coefficients of $S_j$ in the mass eigenstate
$h_{i}\,$, which satisfy $|S_{i1}|^2 + |S_{i2}|^2 + |S_{i3}|^2 =
1$\,, and we assume that $m_{h_1} < m_{h_2} < m_{h_3} $\,. In this
work, we regard $h_2$ as the 125~GeV SM-like Higgs boson, thus
$|S_{23}|^2$ is the singlet component in $h_2$, and
$|S_{23}|^2<0.5$.

\textbf{Neutralino sector:} In the NMSSM there are five neutralinos
($\chi^0_{i}$), which are mixtures of bino ($\tilde{B}$),
wino ($\tilde{W^{0}}$), higgsino ($\tilde{H_{u}}$,
$\tilde{H_{d}}$) and singlino ($\tilde{S}$):
\begin{eqnarray}
\left(  \begin{array}{c}
    \chi^0_{\rm 1} \\
    \chi^0_{\rm 2} \\
    \chi^0_{\rm 3} \\
    \chi^0_{\rm 4} \\
    \chi^0_{\rm 5} \\
  \end{array} \right)
=N_{ij} \left(  \begin{array}{c}
    \tilde{B} \\
    \tilde{W^{0}} \\
    \tilde{H_{u}} \\
    \tilde{H_{d}} \\
    \tilde{S} \\
  \end{array} \right).
\end{eqnarray}
We assume that the lightest neutralino is the lightest SUSY
particle (LSP) and makes up dark matter.

In the basis $\{\,\tilde{B},\, \tilde{W^{0}},\, \tilde{H_{u}},\,
\tilde{H_{d}},\, \tilde{S}\,\}$, the tree-level neutralino mass
matrix takes the form \cite{NMHdecay, NMSSM-rev}
\begin{eqnarray}
M_{\tilde{\chi}^{0}}= \left( \begin{array}{ccccc}
    M_{1} & 0     & \dfrac{g_{1}v_{u}}{\sqrt{2}} & -\dfrac{g_{1}v_{d}}{\sqrt{2}}    & 0 \\[4mm]
    0     & M_{2} & -\dfrac{g_{2}v_{u}}{\sqrt{2}} & \dfrac{g_{2}v_{d}}{\sqrt{2}}    & 0 \\[4mm]
 \dfrac{g_{1}v_{u}}{\sqrt{2}} & -\dfrac{g_{2}v_{u}}{\sqrt{2}}     & 0 & -\mu_{\rm eff} & -\lambda v_{d} \\[4mm]
 -\dfrac{g_{1}v_{d}}{\sqrt{2}} & \dfrac{g_{1}v_{d}}{\sqrt{2}}   & -\mu_{\rm eff} & 0 & -\lambda v_{u} \\[4mm]
    0     & 0                 & -\lambda v_{d}             & -\lambda v_{u} & 2\kappa v_{s}\\
  \end{array} \right).\label{Neu-Matric}
\end{eqnarray}

\textbf{Chargino sector: } The charged higgsinos $\tilde{H}_{u}^+$,
$\tilde{H}_{d}^-$ (with mass scale around $\mu_{\rm eff}$) and the
charged gaugino $\tilde{W}^{\pm}$ (with mass scale $M_2$) can also
mix respectively, forming two couples of physical charginos
$\chi_1^{\pm}, \, \chi_2^{\pm}$.

\textbf{Gluino sector: }
As a gauge boson, each gluon also has a same-color superpartner, which is also sorted into gauginos, and whose mass is close to its soft mass $M_3$. 

\textbf{Squark and slepton sector: } Each  quark or charged
lepton has two chiral-eigenstate superpartners $\tilde{f}_{ L}$ and
$\tilde{f}_{ R}$, which mix to form two mass-eigenstate superpartners.
The mass difference between the two mass eigenstates is
proportional to the corresponding trilinear couplings $A_{ f}$. Since
the Yukawa couplings of the first two generations of fermions are very
weak, the two superpartners of each  fermion can be seen as
mass-degenerate. In the NMSSM, with  only the left-hand state, each
neutrino has only one superpartner, whose mass is equal or close
to its soft mass $m_{\tilde{l}}$.

\textbf{scNMSSM: } In the fully constrained NMSSM
(cNMSSM), like the fully constrained MSSM (cMSSM/mSUGRA), the
soft SUSY breaking terms in the Higgs sector are  assumed to
be unified with those of the squark and slepton sectors at
the GUT scale. However, in the semi-constrained NMSSM (scNMSSM), we
allow the soft SUSY breaking terms in the Higgs sector to be
different. So, in the scNMSSM at the GUT scale, the universal
parameters are \cite{125-scNMSSM, scNMSSM-1405}:
\begin{eqnarray}
\label{eq:gut-universal}
M_1=M_2=M_3 &\equiv& M_{1/2}\, ,\nonumber\\
M^2_{\tilde{q}_{i}}=M^2_{\tilde{u}_{i}}=M^2_{\tilde{d}_{i}}=M^2_{l_{i}}=M^2_{\tilde{e}_{i}} &\equiv& M^2_0 \, , \nonumber\\
A_{t}=A_{b}=A_\tau &\equiv& A_0 \, .
\end{eqnarray}
The Higgs soft masses $M^2_{H_{u}}$, $M^2_{H_{d}}$ and $M^2_{S}$
are allowed to be different from $M^2_0$, and the trilinear
couplings $A_\lambda$, $A_\kappa$ can be different from $A_0$.
Since we have three minimisation equations for the VEVs
\cite{NMSSM-rv0910}, the three Higgs soft masses can be determined
with other parameters. Hence, in the scNMSSM we choose the complete
parameter sector as:
\begin{equation}
\label{eq:parameter} \lambda, \; \kappa, \; \tan\beta,\; \mu_{\rm
eff}, \; A_\lambda, \; A_\kappa, \; A_0, \; M_0, \; M_{1/2}\,.
\end{equation}

\textbf{Parameters running in scNMSSM: } Parameters at the GUT scale
should run via renormalization group equations (RGEs) to
SUSY-breaking scale $M_{\rm SUSY}$. The GUT scale is usually about
$10^{16}$ GeV, and $M_{\rm SUSY}$ is usually chosen to be at
$10^3$ GeV scale. Then, the approximate running of some parameters
can be written as \cite{nmssmtools, NMSSM-RGE}:
\begin{eqnarray}
\label{running}
&& R \equiv (1+\tan^2\!\beta)/(1.29\,\tan^2\!\beta),
      \nonumber\\
&& K \equiv (1-R)\,(A_0-2.24\,M_{1/2})^2+7.84 \, M_{1/2}^2,
      \nonumber\\
&& A_{t} \approx A_0-R\,(A_0-2.24\,M_{1/2})-3.97M_{1/2},
      \nonumber\\
&& A_{b} \approx A_0-R/6\cdot\!(A_0-2.24 M_{1/2})-3.93 M_{1/2},
      \nonumber\\
&& M_{\tilde{q}_3}^2 \approx (1-R/2)\,M_0^2+7.02\,M_{1/2}^2-K\!\cdot\!R/6,
      \nonumber\\
&& M_{\tilde{u}_3}^2 \approx (1-R)\,M_0^2+6.6\,M_{1/2}^2-K\!\cdot\!R/3,
      \nonumber\\
&& M_{\tilde{d}_3}^2 \approx M_0^2+6.55 M_{1/2}^2,
      \qquad 
M_{\tilde{q}_2}^2 \approx M_0^2 +7.02 M_{1/2}^2,
      \nonumber\\
&& M_{\tilde{u}_2}^2 \approx M_0^2 +6.6 M_{1/2}^2,
      \qquad~ 
M_{\tilde{d}_2}^2 \approx M_0^2 +6.55 M_{1/2}^2,
      \nonumber\\
&& A_{\tau} \approx A_0-0.69 M_{1/2},
      \qquad~~ 
      M_{\tilde{\ell}_3}^2 \approx M_0^2+0.52 M_{1/2}^2,
      \nonumber\\
&& M_{\tilde{e}_3}^2 \approx M_0^2+0.15 M_{1/2}^2,
      \qquad 
      M_{\tilde{\ell}_2}^2 \approx M_0^2 +0.52 M_{1/2}^2,
      \nonumber\\
&& M_{\tilde{e}_2}^2 \approx M_0^2 +0.15 M_{1/2}^2,
      \qquad~ 
      M_1 \approx 0.4 M_{1/2},
      \nonumber\\
&& M_2 \approx 0.8 M_{1/2},
      \qquad\qquad\quad~~ 
      M_3 \approx 2.4 M_{1/2}.
\end{eqnarray}

\section{Numerical results and discussion}

In this work, we use the program $\textsf{NMSPEC\_MCMC}$
\cite{nmspec} in $\textsf{NMSSMTools\_5.2.0}$
\cite{nmssmtools} to scan the parameter space of the scNMSSM by
considering various experimental constraints. We chose the
parameter space to scan as follows:
\begin{eqnarray}
\label{eq:space} 0.3<\lambda<0.7, & 0<\kappa<0.7, &  1<\tan\beta <30, \nonumber\\[2mm]
100<\mu_{\rm eff}<200 \GeV,       &  0<M_0<500 \GeV,   &  0<M_{1/2}<2 \TeV, \nonumber\\[2mm]
|A_0|<10 \TeV, & |A_\lambda |<10 \TeV, & \quad |A_\kappa |<10 \TeV,\nonumber\\
\end{eqnarray}
where we have the following considerations in our choice of parameter space:

 1) Small $\mu_{\rm eff}$ and $M_0$, to get large muon g-2 and also low fine tuning.

 2) Large $\lambda$ ($>0.3$) to make our results much different from those of the MSSM in Higgs physics, since there is only one term different in the superpotential between NMSSM and MSSM: $\lambda\hat{S}\hat{H}_{u}\hat{H}_{d}$
 in Eq.~(\ref{eq:w_nmssm}), for the doublet-singlet mixing.

 3) Smaller $\tan\beta$ ($<30$) than in MSSM, as in the NMSSM scenario of $h_2$ as the 125 GeV SM-like Higgs. In this scenario, we should have $|M_{23}|\ll |M_{33}|<M_{22}$ in the Higgs mass matrix Eq.~(\ref{eq:m}), thus
      \begin{eqnarray}
      M_A\approx \dfrac{2\mu}{\sin2\beta}\approx \mu\tan\beta \approx A_{\lambda}+\frac{\kappa}{\lambda}\mu.
      \end{eqnarray}
       $A_\lambda$ at the SUSY breaking scale should not be too large, since we have another term of doublet-singlet mixing $\lambda A_\lambda S H_{u}.H_{d}$ in the soft breaking terms in Eq.~(\ref{eq:higgssoft}).

In the scan, we required the surviving samples to satisfy the following
constraints:

  1) Theoretical constraints of vacuum stability, and no Landau pole in running $\lambda$, $\kappa$, and Yukawa couplings below $M_{\rm GUT}$ \cite{nmssmtools, nmspec}.

  2) The second light scalar CP-even Higgs, $h_2$, as the SM-like Higgs boson with mass around $125\GeV$ (e.g., $123\!<\!m_{h_2}\!<\!127 \GeV$), with its production rates fitting LHC data globally.
      For the global fit we used a method like that in our former works \cite{low-NMSSM, 1311-mdm},  with the Higgs data updated with Fig.~3 from Ref.~\cite{15007-Atlas} and the left part of Fig.~5 from Ref.~\cite{1412-CMS}.
      There are 20 experimental data sets in total, so we require $\chi^2\leq31.4$, which means each surviving sample fits 20 experimental data sets at $95\%$ confidence level.

  3) Constraints of searches for low mass and high mass resonances at LEP, Tevatron, and LHC. These constrain the production rates of light and heavy Higgs.
      We implemented these constraints by the package \textsf{HiggsBounds-5.1.1beta} \cite{higgsbounds}.
      We also required the mass of the light Higgs to be $65\!\sim\! 122\GeV$, since we checked that below $65\GeV$ its diphoton rate is always very small because of the strong constraints at LEP.
      Also, when the light Higgs is lighter than 62 GeV, exotic decays of the 125 GeV Higgs will be generated, which we have discussed in detail in our former paper \cite{low-NMSSM}.

  4) Constraints of searches for squarks of the first two generations and gluinos at Run I of the LHC$^{1)}$.\
  \footnotetext{1) For the stop mass, we checked that our result satisfies the simulation result of $m_{\tilde{t}_1}\gtrsim 500\GeV$ in Ref.~\cite{nNMSSM-MC}.
      For 13 TeV search results at the LHC, all these bounds may be a little higher, but we checked that with stricter constraints, e.g.,
      $m_{\tilde{q}_{1,2}}>1200\GeV,\; m_{\tilde{g}}>1800\GeV$, and $m_{\tilde{t}_1}> 600\GeV$, our results, such as muon g-2, do not change much.
      We will check the exact bounds of these sparticle masses in this model in our future work by doing detailed simulations. }
We follow the result in Ref.~\cite{scNMSSM-1405}:
    \begin{equation}
        \label{eq:squarkmass}
        m_{\tilde{q}_{1,2}}\gtrsim900\GeV, \quad m_{\tilde{g}}\gtrsim1400\GeV \,.
    \end{equation}
    We use the constraints of mass bounds of chargino and sleptons from LEP.
      We also checked our surviving samples with \textsf{SModelS-v1.1.1} \cite{smodels1} (including database v1.1.2 \cite{sms-db112})$^{2)}$.
\footnotetext{2) This includes many constraints on stop $\tilde{t}_1$ \cite{Sirunyan:2017cwe, Sirunyan:2017kqq, Sirunyan:2017leh, Sirunyan:2017xse},
chargino $\chi^\pm_1$ and neutralino $\chi^0_2$ \cite{Sirunyan:2017zss, Sirunyan:2018ubx}. We checked that it cannot give the surviving
samples further strong constraints, because for the surviving samples: $\chi^\pm_1$ are Higgsino-like, and $\chi^0_{1,2}$ are Higgsino or singlino-like,
thus light $\chi^\pm_1$ and $\chi^0_2$ mainly decay to $\chi^0_1$ and a pair of quarks, each channel with about 10-20 percent; most charginos and
neutralinos are lighter than $\tilde{t}_1$. Thus $\tilde{t}_1$ can have many decay channels, where even the dominant channel, e.g., $\tilde{t}_1\to b \chi^+_1$, cannot be over half.}

  5) Constraints from B physics, such as $B_{s} \to \mu^+ \mu^-$, $B_{d} \to \mu^+ \mu^-$, $B \to X_{s} \gamma$ and $ B^+ \to \tau^+ \nu_\tau$, etc.
      \cite{BaBar-Bph, LHCb-BsMuMu, PDG2016}.
      \begin{eqnarray}
      && 1.7\times 10^{-9} < Br( B_{s} \to \mu^+\mu^-) < 4.5\times 10^{-9},\nonumber\\
      && 1.1\times 10^{-10} < Br( B_{d} \to \mu^+\mu^-) < 7.1\times 10^{-10}, \nonumber\\
      && 2.99\times 10^{-4} < Br( B \to X_{s} \gamma) < 3.87\times 10^{-4},\nonumber\\
      && 0.70\times 10^{-4} < Br( B^+ \to \tau^+ + \nu_\tau) < 1.58\times 10^{-4}.
\end{eqnarray}

  6) Constraints from dark matter relic density from WMAP/Planck \cite{dm-omg, PDG2016}, the spin-independent (SI)
  results of direct searches for dark matter at LUX 2017 \cite{LUX2017}, PandaX-II 2017 \cite{PandaXII2017}, and XENON1T 2018 \cite{XENON1T2018},
      and the spin-dependent (SD) results of direct searches for dark matter by PICO, LUX, and PandaX-II in 2016 \cite{dm-sd2016}.
      We require the lightest neutralino $\chi^0_1$ to be the dark matter candidate.
      For the relic density, we only apply the upper bound, e.g., $0\leqslant \varOmega \leqslant 0.131$,  considering that there may be other sources of dark matter \cite{low-NMSSM98, nmssmtools}.

  7) The constraint of the muon anomalous magnetic moment (muon g-2) at 2$\sigma$ level including the theoretical error. For the experimental data and SM calculation  without boson contributions, we use \cite{mug2-ex2006, mug2-sm}:
      \begin{eqnarray}
        a_\mu^{\rm ex} &=& (11659208.0 \pm 6.3) \times 10^{-10}, \\[2mm]
        \delta a_\mu &\equiv& a_\mu^{\rm ex}-a_\mu^{\rm SM} = (27.4\pm9.3) \times 10^{-10}
      \end{eqnarray}
      We calculate the SUSY contribution $\delta a_\mu$ including SM-like bosons, and require it to satisfy $\delta a_\mu$ at $2\sigma$ level.
       We also include our error in the SUSY $\delta a_\mu$ calculation, which is about $1.5 \times 10^{-10}$.

  8) The theoretical constraint of low fine tuning from the GUT scale, which is defined by \cite{NMSSM-FT}:
      \begin{equation}
      \label{eq:FT}
      FT={\rm Max}\left\{ \left|\frac{\partial\, {\rm ln}(M_{Z})}{\partial\, {\rm ln}(p^{\rm GUT}_{i})}\right| \right\} \, ,
      \end{equation}
      where each $p^{\rm GUT}_{i}$ denotes a parameter at the GUT scale:
      \begin{equation}
      p^{\rm GUT}_{i} = M_{\tilde{H}_{u}}, M_{\tilde{H}_{d}}, M_{\tilde{S}}, M_0, M_{1/2},
      A_{\lambda}, A_{\kappa}, A_0, \lambda, \kappa, y_{t}, g, M_{\rm GUT},
      \end{equation}
      where $g=\sqrt{g_1^2+g_2^2}/2$, $y_{t}$ is the Yukawa coupling of the top quark, and $M_{\rm GUT}$ is the GUT scale.
      We require $FT<1000$ for each surviving sample.

We take a modified multi-path Markov Chain Monte Carlo (MCMC) scan
in parameter space, where we do not use likelihood functions.
Instead we require each good point to satisfy all our experimental
constraints at $2\sigma$ level, or below the upper limits of
$95\%$ (for the DM direct detection, it is $90\%$ following data
released by the collaborations). Each time we get a good point
surviving all our constraints, we save the point, and search for the
next good point around the former one. Since we use a Gaussian
random number and set a not-small standard step, the later good
point can be much different from the
former one, which ensures we get as much as possible of the surviving parameter space
available. In total, we get nearly $10^6$ surviving samples. As
some samples may be very similar to each other, we remove most
repetitive samples by calculating the distance between them.
First, we normalize all samples by using min-max normalization
(MMN), which is just a linear transformation of the original data.
We normalize each sample to 9 dimensions, as there are 9 free
parameters $x^i$ in the scan
\begin{equation}
\label{eq:norm}
\hat{x}^i=\frac{x^i-x^i_{\rm min}}{x^i_{\rm max}-x^i_{\rm min}} \qquad
(i=1,...,9)
\end{equation}
After this linear transformation, all of these 9 new parameters
$\hat{x}^i$ will fall in $[0,1]$. Then we calculate the Euclidean
distance between all these surviving samples. If the distance
between two points is too small, we just select one of them
randomly. For each panel in the following figures, to make them
look good and be of small size, we take a MMN similarly but in 3
dimensions, which are the horizontal, vertical and color-indicated
quantities.

\begin{figure*}
  \centering
\includegraphics[width=16cm]{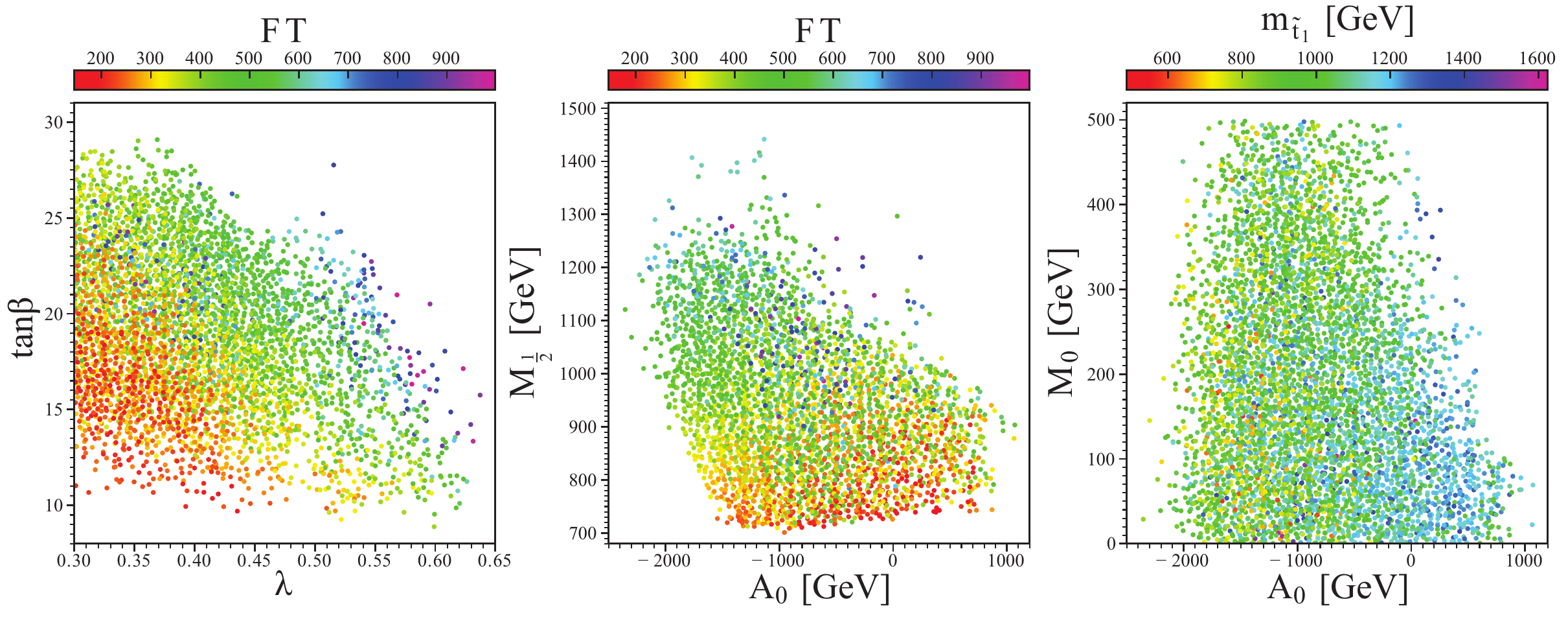}
\figcaption{\label{fig1}(color online) Surviving samples in the $tan\beta$ versus $\lambda$
(left), $M_{1/2}$ versus $A_0$ (middle), and $M_0$ versus $A_0$
(right) planes. Colors in the left and middle panels indicate fine
tuning from GUT scale, while colors in the right panel indicate
the mass of the lighter stop $\tilde{t}_1$. }
\end{figure*}

In Fig.~\ref{fig1}, we project the surviving samples on
the $\lambda-\tan\beta$, $A_0-M_{1/2}$ and $A_0-M_0$ planes.
We show fine tuning from the GUT scale (left and
middle panels) and the lighter stop mass $m_{\tilde{t}_1}$ (right
panel) by different colors.
We can see from the left and middle panels that fine tuning $FT$ can be as low as around 150 at most. 
In the left panel, we can also see that low-fine-tuning samples are
mostly located in the $\tan\beta\lesssim 15$, or
$15\lesssim\tan\beta\lesssim 25$ but $\lambda\lesssim0.4$ regions.
This is because, according to the minimisation equation of $v_{u}$
\cite{NMSSM-FT},
\begin{eqnarray}
M_{H_{u}}^2+\mu^2+\frac{1}{2}X m_Z^2=0,
\end{eqnarray}
where
\begin{eqnarray}
X &=&  \frac{\tan^2\beta-1}{\tan^2\beta+1}+\frac{\tan^2\beta}{\tan^2\beta+1}
\frac{3y_{t}^4}{8\uppi^2 g^2} \ln{\frac{{m^2_{\tilde{t}}}}{m_{t}^2}} \nonumber\\
&& -\frac{1}{\tan\beta} \left(\frac{\mu A_\lambda}{m_Z^2}+\frac{\kappa}{\lambda}\frac{\mu^2}{m_Z^2}\right)
+\frac{\lambda^2}{g^2}\frac{2}{\tan^2\beta+1}
\end{eqnarray}
is a function of $\lambda,\,\tan\beta$, etc. We checked that for
most of the surviving samples, the largest fine tuning comes from
parameter $M_{H_{u}}$,
and that of the rest comes from parameter $\lambda$. According to
RGE running, $|M^2_{H_{u}}|$ is related to $M_0,\, M_{1/2},\,
{\rm and}\, A_0$, thus we can see from the middle panel that
samples with small $M_{1/2}$ and $A_0$ usually have low fine
tuning. From the middle and right panels, we can see that these
surviving regions are not symmetric around $A_0=0$, where negative
$A_0$ is more favored. This is because at SUSY-breaking scale we
have $M_{\tilde{q}_3}^2 \approx M_0^2+6.55 M_{1/2}^2$ and
\begin{eqnarray}
A_{t}  &\approx&   A_0- \frac{(A_0-2.24M_{1/2})(1+\tan^2 \beta)}{(1.29\tan^2\!\beta)}-3.97M_{1/2} \nonumber\\
&\approx&   0.22(A_0-10M_{1/2})
\end{eqnarray}
according to the RGEs. We have $M_{1/2}\gtrsim 700 \GeV$, mainly
because we require $m_{\tilde{g}} \gtrsim 1400 \GeV$, while at
SUSY-breaking scale $M_3\approx 2.4 M_{1/2}$. Later we
can see from Fig.~\ref{fig4} that $M_{1/2}$ has upper bounds of
about $1500\GeV$ mainly because of the constraint of muon g-2.
Finally, in the right panel, we can see the mass of the stop can be as
low as about 500 GeV. We will continue studying these light-stop
cases in our future work, by doing detailed simulations based on the
search results at the 13 TeV LHC.

\begin{figure*}
  \centering
\includegraphics[width=16cm]{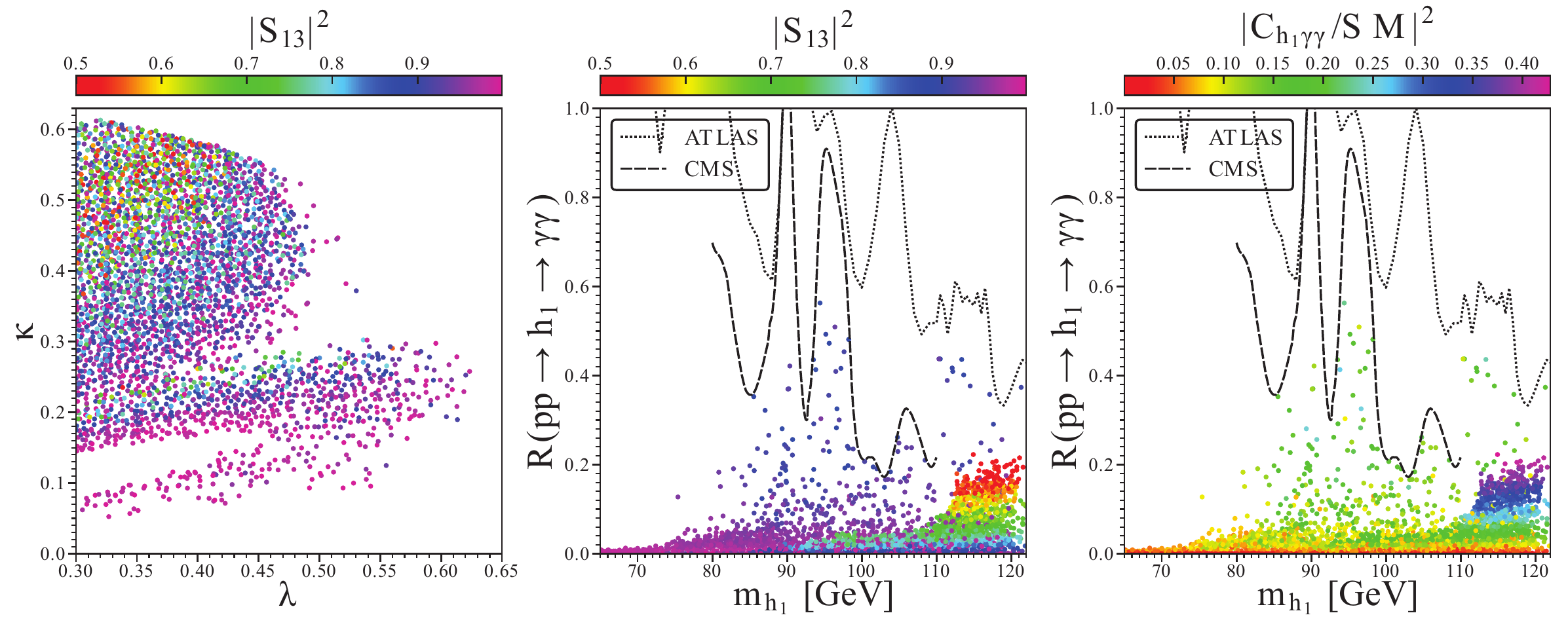}
\figcaption{\label{fig2}(color online) Surviving samples in the $\kappa$ versus $\lambda$
(left), and diphoton production rate of the lightest Higgs $ h_1$
versus its mass $m_{ h_1}$ (middle and right) planes. In the middle
and right panels, the black dotted and dashed line indicates the
observed exclusion limits (95\% CL) from ATLAS
\cite{Atlas2014-lightH} and CMS \cite{CMS95} on $R( pp\to
h_1\to\gamma\gamma)$ respectively. Colors in the left and middle
panels indicate the singlet component in $h_1$, while colors in the
right panel indicate the squared effective coupling of $ h_1$ with two
photons, reduced by its corresponding SM value, i.e. $|C_{ h_1
\gamma\gamma} /SM|^2$. }
\end{figure*}

In Fig.~\ref{fig2}, we project the surviving samples on the $\lambda$
versus $\kappa$ (left), and $R( pp\to h_1\to \gamma\gamma)$ versus
$m_{ h_1}$ (middle and right) planes respectively. We show the
singlet component in $ h_1$ (left and middle panels) and the
reduced squared coupling $|C_{ h_1 \gamma\gamma}/SM|^2$ (right
panel) by different colors. We can see from the left and middle
panels that most of the samples have $|S_{13}|^2$ approaching 1,
which means they are highly singlet-dominated. The singlet
component in $ h_1$ is 0.5 at least, since $ h_2$ is the SM-like
Higgs. It can be sorted into two regions in the $\lambda-\kappa$
plane:
\begin{eqnarray}
& \lambda\gtrsim1.5\kappa {\rm ~region} , {\rm where} & h_1 {\rm ~are~highly-singlet-dominated~} \nonumber\\
& &  (|S_{13}|^2\gtrsim0.8) \nonumber\\
& \lambda\lesssim1.5\kappa {\rm ~region}, & {\rm including~smaller-}|S_{13}|^2 {\rm ~samples}\nonumber\\
& & (0.5 \lesssim |S_{13}|^2\lesssim 0.8)
\end{eqnarray}
The samples of $|S_{13}|^2\lesssim 0.9$ are mainly in the latter
region, because in the former region, with small $\kappa/\lambda$
we will have $|M_{23}^2|\ll |M_{22}^2|$, which will result in very
little mixing between singlet and SM-like doublet, thus very
little singlet component in $ h_2$ and very little doublet
component in $ h_1$. From the middle panel, we  find that
doublet-singlet mixing can only be considerable
($|S_{13}|^2\lesssim0.9$) when $m_{ h_1}\gtrsim90\GeV$. Combining
the middle and right panels we can also see that some
highly-singlet-dominated $h_1$ samples ($|S_{13}|^2\gtrsim 0.8$)
can provide a considerable diphoton rate $R( pp\to h_1\to
\gamma\gamma)$, while the rates are not so large for
smaller-$|S_{13}|^2$ samples ($0.5 \lesssim |S_{13}|^2\lesssim
0.8$). This is because for the former samples, we have light
higgsino-like chargino (see Fig.~\ref{fig4}) and moderate
$\lambda$, thus large $ h_1 \gamma\gamma$ loop-reduced coupling and
large $ h_1 \to \gamma\gamma$ branching ratio. For the latter
samples, the $ h_1$ reduced coupling to $ \gamma\gamma$ can be smaller
than to other SM particles like $ b\bar{b}$, thus the $ h_1 \to
\gamma\gamma$ branching ratio cannot be large. We checked that the
reduced $ h_1\gamma\gamma$ coupling can be two times that of the doublet
component in $ h_1$ ($1-|S_{13}|^2$) for the former, while it can
only be about 0.5 for the latter. According to the latest result of
the search for low-mass resonances by CMS, the suspected resonance
is at around 95 GeV, with a diphoton rate of about $0.5\pm 0.2$
\cite{CMS95}. We can see that we have some samples providing such
a signal. In Table~\ref{table1}, we provide the detailed information
of four such samples for further study. The search results for
low-mass resonances by ATLAS at Run I of the LHC
\cite{Atlas2014-lightH} are also shown on the middle and right
panels. We can see that the upper limit from ATLAS is higher than that from
CMS, and further results from ATLAS are needed to cross check the
suspected excess.

\begin{table*}[hp]
\tabcaption{\label{table1}Four representative samples predicting the
 diphoton rate hinted at by CMS data, where
$R_{ h_1\to\gamma\gamma}$ is the same as $R( gg\to
h_1\to\gamma\gamma)$ elsewhere in this paper.}
\footnotesize
\vspace{-1mm}
\begin{tabular*}{176mm}{c@{\extracolsep{\fill}}ccccc ccccc cc}
\toprule

& $ $ & $\lambda$ & $\kappa$ & $\tan\beta$  & $\mu_{\rm eff}$/GeV & $A_\lambda^{\rm GUT}$/GeV  & $A_\kappa^{\rm GUT}$/GeV  & $A_0$/GeV & $M_0$/GeV  & $M_{1/2}$/GeV  & $m_{ h_1}$/GeV & $R_{ h_1\to\gamma\gamma}$  \\
\hline
& P1 & 0.51 & 0.26 & 20.2 & 130.3 & 3856.1 & 2707.1 & -172.6 & 23.3 & 894.3 & 96.7 & 0.33 \\
& P2 & 0.42 & 0.22 & 16.1 & 145.5 & 2271.8 & 953.5 & -997.1 & 163.3 & 754.0 & 95.6 & 0.47 \\
& P3 & 0.31 & 0.18 & 21.4 & 165.8 & 2923.6 & 439.1 & -1084.2 & 10.8 & 1006.2 & 96.4 & 0.51 \\
& P4 & 0.37 & 0.20 & 18.7 & 137.8 & 2363.8 & 700.0 & -826.6 & 97.5 & 827.9 & 95.2 & 0.49 \\
\bottomrule
\end{tabular*}
\end{table*}

\begin{figure*}[hp]
  \centering
\includegraphics[width=16cm]{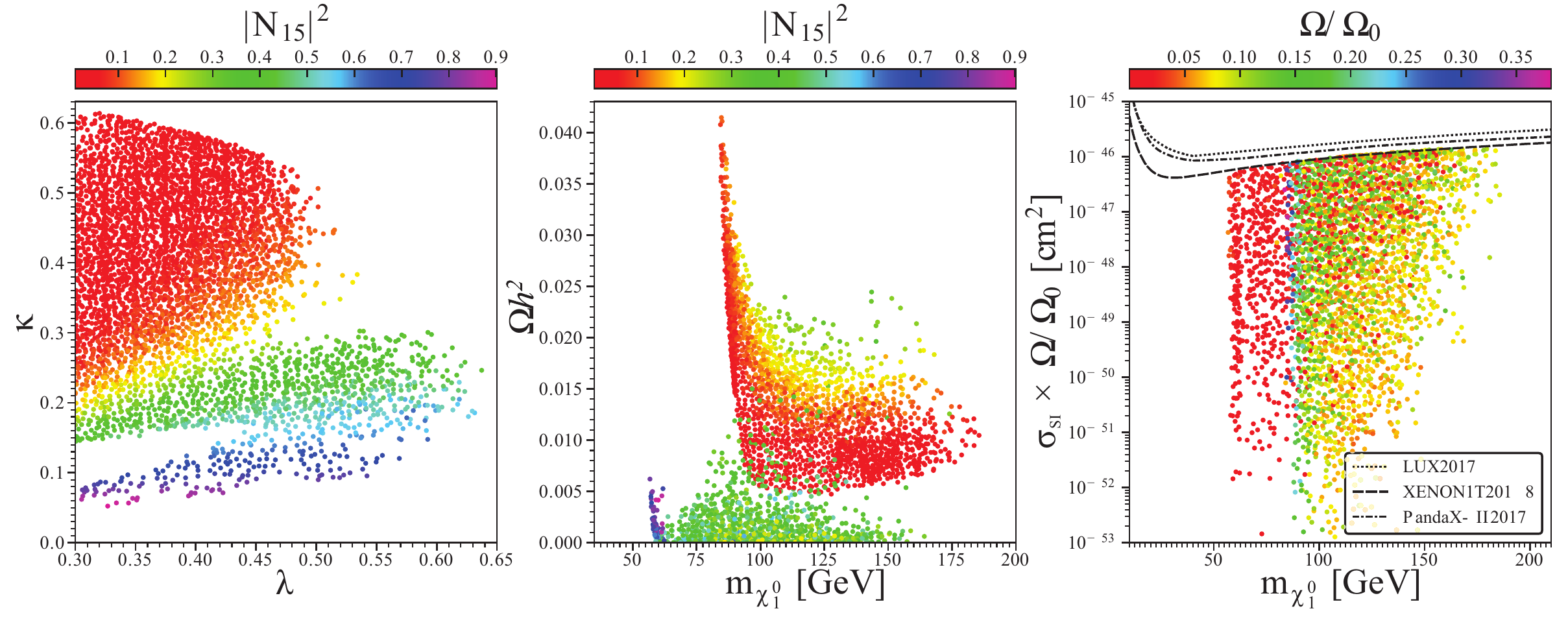}
\caption{Surviving samples in the $\kappa$ versus $\lambda$
(left), dark matter relic density $\varOmega h^2$ versus the lightest
neutralino (LSP) mass $m_{\chi^0_1}$ (middle), and spin-independent
dark matter and nucleon scattering cross section
($\sigma_{\mbox{\rm\tiny SI}}\!\!\times\!\varOmega/\varOmega_0$) versus
LSP mass $m_{\chi^0_1}$ (right) planes. In the middle and right
panels, the black dotted, dot-dashed, and dashed lines indicate
the observed exclusion limits (90\% CL) on $\sigma_{\mbox{\rm\tiny
SI}}\!\!\times\!\varOmega/\varOmega_0$ released by LUX 2017, PandaX-II
2017 and XENON1T 2018, respectively. Colors in the left and middle
panels indicate the singlino component in $\chi^0_1$, while colors in
the right panel indicate the ratio of LSP relic density in the
observed value ($\varOmega/\varOmega_0$), where $\varOmega_0 h^2=0.1187$
\cite{PDG2016, dm-omg}. } \label{fig3}
\end{figure*}

\begin{figure*}
  \centering
\includegraphics[width=16cm]{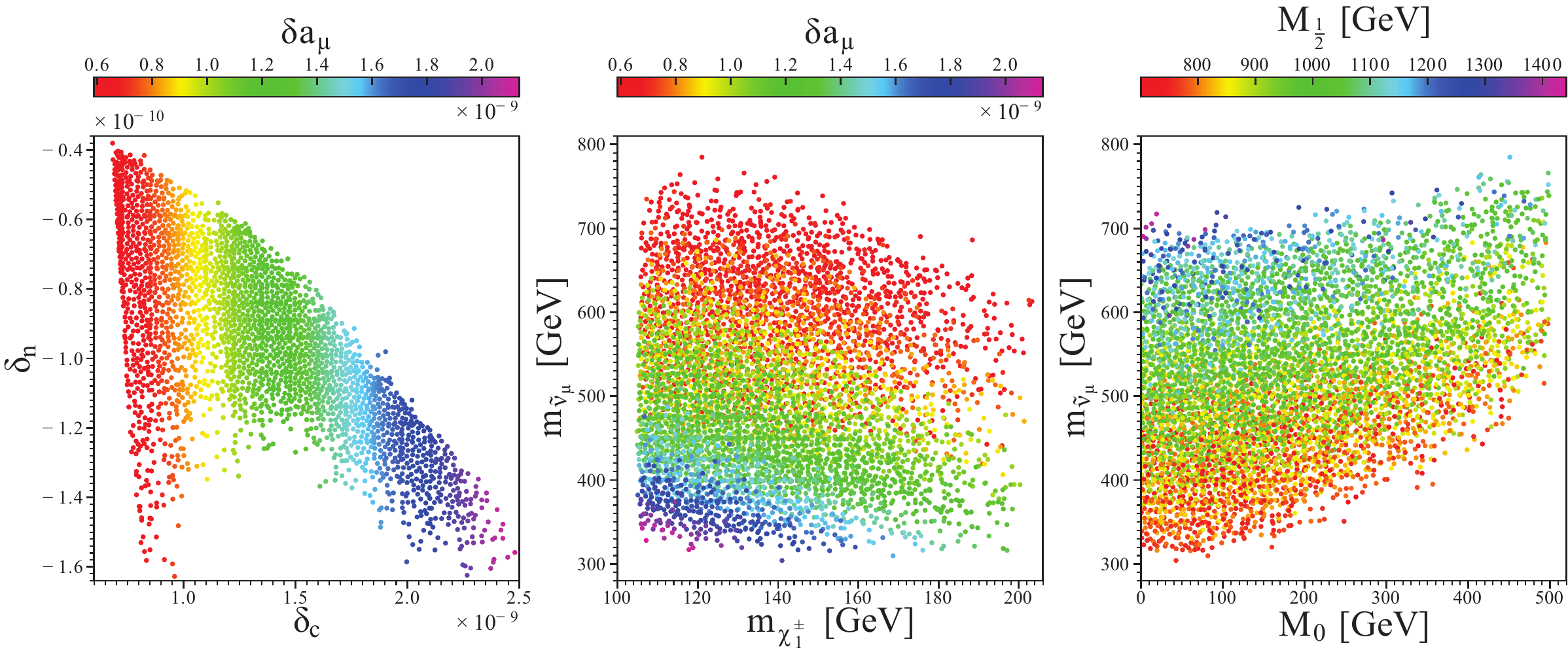}
\figcaption{\label{fig4}(color online) Surviving samples in the neutralino-smuon contribution
$\delta_n$ versus chargino-sneutrino contribution $\delta_c$ to
muon g-2 (left), light chargino mass $m_{\chi^\pm_1}$ versus muon
sneutrino mass $m_{\tilde{\nu}_\mu}$ (middle), and
$m_{\tilde{\nu}_\mu}$ versus parameter $M_0$ (right) planes.
Colors in the left and middle panels indicate muon g-2 ($\delta
a_{\mu}$), while colors in the right panel indicate parameter
$M_{1/2}$. }
\end{figure*}

In Fig.~\ref{fig3} we show the properties of dark matter in the
scNMSSM. In this work, we require the lightest neutralino
$\chi^0_1$ to be the LSP and to constitute dark matter by a ratio
of $\varOmega/\varOmega_0$, with the right relic density $\varOmega_0
h^2=0.1187$ \cite{PDG2016, dm-omg}. Hence we adjust the SI
scattering cross section of each sample by tuning the
corresponding ratio $\varOmega/\varOmega_0$. Since the results of
searches for gluinos at the LHC require $M_3\simeq
m_{\tilde{g}}\gtrsim1400\GeV$, and the universal gaugino mass at
GUT scale requires
\begin{eqnarray}
M_1:M_2:M_3\approx1:2:6,
\end{eqnarray}
while $\mu_{\rm eff}\!<\!200\GeV$ and $\kappa v_S =
\mu_{\mbox{\rm\tiny eff}}\!\cdot\! \kappa/\lambda$, according to
the tree-level neutralino mass matrix Eq.(\ref{Neu-Matric}), we
can infer that the main components of $\chi^0_1$ can only be the
singlino and higgsino. From the left panel of Fig.\ref{fig3}, we
can see clearly that
\begin{eqnarray}
{\rm when~} \lambda\gtrsim1.5\kappa: & |N_{15}|^2\gtrsim0.3,&
\chi^0_1 {\rm ~is~singlino-dominated}
\nonumber\\
{\rm when~} \lambda\lesssim1.5\kappa:& |N_{15}|^2\lesssim0.3,&
\chi^0_1 {\rm ~is~Higgsino-dominated.}\nonumber\\
\end{eqnarray}
We can categorize the surviving samples into three classes, which
can be called the $h/Z$ funnel, focus point, and $ A_1$ funnel scenarios
respectively, as in Ref.~\cite{susy-dm}.

 1) From the middle and right panels of Fig.~\ref{fig3}, we can see that in the $h/Z$ funnel scenario, its mass $m_{\chi^0_1}\lesssim m_{ h_2}/2$, and its relic density is only about $1/10$ of
 the WMAP data at most. For some samples the SI scattering cross section before adjustment with $\varOmega/\varOmega_0$ are above the exclusion limit by XENON1T 2018, LUX 2017, and PandaX-II 2017.
      Combining with the left panel, we can see that, in the $h/Z$ funnel scenario, the larger $\lambda$ the smaller its relic density.
      This is because a pair of singlet-dominated $\chi^0_1$ annihilates to a pair of SM particles through the SM-like Higgs $ h_2$, and the coupling $ h_2 \chi^0_1 \chi^0_1$ is proportional to $\lambda$:
      \begin{eqnarray}
      C_{ h_2\chi^0_1\chi^0_1}&=&\sqrt{2}\lambda N_{15} \big[ S_{21}(N_{14}\cos\beta-N_{13}\sin\beta)\nonumber\\
      && +S_{22}(N_{14}\sin\beta+N_{13}\cos\beta) \big] -\sqrt{2}\kappa S_{23}|N_{15}|^2 \nonumber\\
      && +(g_2 N_{12}-g_1 N_{11}) \big[ S_{21}(N_{13}\cos\beta+N_{14}\sin\beta)\nonumber\\
      && +S_{22}(N_{13}\sin\beta-N_{14}\cos\beta)\big] \nonumber\\
      &\approx& \sqrt{2}\lambda S_{22}N_{15}(N_{14}\sin\beta+N_{13}\cos\beta) \nonumber\\
      &&     -\sqrt{2}\kappa S_{23}|N_{15}|^2,
      \end{eqnarray}
      since $|S_{22}|\gg |S_{23}| \gg |S_{21}|$,\;
      $|N_{15}|\gg |N_{13,14}| \gg |N_{11,12}|$,\; and $\lambda>\kappa$.
      Besides, the SI scattering of $\chi^0_1$ with SM particles is also mainly mediated by the SM-like $ h_2$.
      Thus, we can infer that, with smaller $\lambda$, the relic density of $\chi^0_1$ can be larger, and the SI scattering cross section can be smaller.
      Of course, to have $\chi^0_1$ singlino-dominated, we also need even smaller $\kappa$.

 2) In the focus point scenario, when $m_{\chi_1^0}$ is slightly larger than $m_{ W}$, the main annihilation mechanism is $\chi^0_1\chi^0_1 \to W^+W^-$ though the t or u channel chargino, or s channel $Z$ or scalars.
      A peak of relic density appears around $m_{\chi_1^0} \thickapprox m_{ W}$ in the middle panel of Fig.~3, because the relic density is inversely proportional to $\sqrt{1-m_{ W}^2/m_{\chi^0_1}^2}$ \cite{susy-dm-1995}.
      The relic density cannot be even larger, because it is very hard for a higgsino-like $\chi^0_1$ to unlimitedly approximate to $m_{ W}$, and also the results for SI scattering cross section in
      2018 give even stronger constraints on this scenario.

 3) For the other samples, including both singlet-dominated and higgsino-dominated $\chi^0_1$ cases, the main annihilation mechanism is the $ A_1$ funnel, where the light CP-odd scalar $ A_1$
  is usually singlet-dominated ($\gtrsim 90\%$), but has a $\varphi_{d}$ composition of several percent.
      Thus there can be a large $ A_1\chi^0_1\chi^0_1$ coupling for large $\lambda$ or $\kappa$, and a considerable $ A_1b\bar{b}$ coupling for the $\varphi_{d}$ composition and large $\tan\beta$.
      In this scenario, a pair of $\chi^0_1$ mainly annihilates through $ A_1$, and  $b\bar{b}$ and $\tau^+\tau^-$ are produced.

We also consider the spin-dependent (SD) results of direct
detection for dark matter \cite{dm-sd2016}. However, we checked
that the current upper exclusion limits of SD results are much
higher than the SI ones, and they impose no further constraints on
our surviving samples. So in this work, we do not discuss the SD
results further.

From the left panel of Fig.~\ref{fig4}, we can
interpret muon g-2 ($\delta a_{\mu}$) at $1 \sigma$ level, and the
main contribution comes from the loop of chargino $\chi^{\pm}_1$
and muon sneutrino $\tilde{\nu}_{\mu}$. The loop of the lightest
neutralino $\chi^0_1$ and smuon $\tilde{\mu}_{i}$ cannot contribute as
much as in the MSSM, because $\chi^0_1$ is singlino-dominated
or higgsino-dominated, neither of which has a strong enough coupling with the muon and
its partner. From the middle panel of Fig.~\ref{fig4} we can see
that the chargino loop can contribute much because both the chargino
$\chi^0_1$ and muon sneutrino $\tilde{\nu}_{\mu}$ can be very
light. The lighter they are, the larger muon g-2 is. From the
right panel, the sneutrino mass is mainly
determined by $M_0$ and $M_{1/2}$. In fact, the relation is
roughly $m_{\tilde{\nu}_{\mu}} \approx
\sqrt{M_0^2+0.52M_{1/2}^2}$. Combined with $M_3\approx2.4M_{1/2}$,
we can infer and have checked that, with higher gluino mass
$m_{\tilde{g}}\approx 2\TeV$, the sneutrino mass can still be low as
about $400\GeV$, and thus muon g-2 can still satisfy the data at $1\sigma$
level.

\section{Conclusions}

In this work, we have checked the status of the scNMSSM under current
constraints, such as 125 GeV Higgs data, searches for low and high
mass resonances, searches for SUSY particles at the LHC, B
physics, muon g-2, dark matter relic density by WMAP/Planck, and
direct searches for dark matter by LUX 2017, PandaX-II 2017, and
XENON1T 2018. First, we scanned the parameter space of the scNMSSM in 9
dimensions with the MCMC method. For each valid sample, we calculated
its various physical quantities and required them to satisfy
corresponding constraints. For the surviving samples, we analyzed
fine tuning from the GUT scale, SUSY particle masses, the light scalar and
its diphoton signal, dark matter relic density and direct
detection, muon g-2, and their favoured parameter space. Finally,
we come to the following conclusions regarding the scNMSSM:

  1) For low fine tuning samples, small  $\mu_{\rm eff},$ $M_0,$ $M_{1/2}$, $A_0$, are more favored, and the lighter stop mass can be as low as about 500 GeV,
  which can be further checked in future works with search results at the 13~TeV LHC.

  2) For light higgsino-like charginos and moderate $\lambda$, the highly-singlet-dominated light scalar can have a considerable diphoton rate, satisfying  the latest results of the search for low-mass resonances by CMS.

  3) For high gluino bounds at the LHC and the condition of universal gauginos at the GUT scale, the lightest neutralino can only be singlino-dominated or higgsino-dominated.
  Their mass regions are $m_{\chi^0_1}\lesssim m_{ W}$ and $m_{ W} \lesssim m_{\chi^0_1}\lesssim 200 {\rm ~ GeV}$, and annihilation scenarios are mainly $h/Z$ funnel and focus point
  respectively. The results for SI scattering cross section in 2017/2018 give strong constraints,  especially for the focus point scenario.

  4) For light muon sneutrino and light higgsino-like charginos, we can get large muon g-2, while the contribution of neutralinos cannot be large because bino-like and wino-like neutralinos are heavy.

  5) The model can satisfy all the above constraints, although it is not easy for the lightest neutralino, as the only dark matter candidate, to get enough relic density.

Considering the disadvantage of the scNMSSM, one can  try  three main
kinds of ways to raise the relic density:

    1) Considering other source of relic density, e.g., the effects of modifications of the expansion rate and of the entropy content in the early universe.

    2) Changing the LSP to another sparticle, such as bino-like neutralinos in the non-universal gaugino cases, or sneutrinos in the right-handed neutrinos extended case.

    3) Reducing $\lambda$ and $\kappa$ in the $h/Z$ funnel scenario, although this way may lose a light Higgs.

\end{multicols}

\vspace{-2.5mm} \centerline{\rule{80mm}{0.1pt}} \vspace{1mm}

\begin{multicols}{2}

\end{multicols}

\clearpage

\end{document}